\documentstyle[12pt]{article}

\let\ni=\noindent

\begin{document}

\baselineskip 0.73cm

\renewcommand{\thefootnote}{\fnsymbol{footnote}}

\newcommand{\CKM}{Cabibbo---Kobayashi---Maskawa }

\newcommand{\SK}{Super--\-Kamio\-kande }

\pagestyle {empty}

~~~

\vspace{1cm}

{\large \centerline {\bf Are there two sterile neutrinos}}

{\large \centerline {\bf cooscillating with $\nu_e $ and $\nu_\mu $?{\footnote
{Dedicated to Jan {\L}opusza\'{n}ski in honour of his 75th birthday.}}}}

\vspace{1.0cm}

{\centerline {\sc Wojciech Kr\'{o}likowski}}

\vspace{1.0cm}

{\centerline {\it Institute of Theoretical Physics, Warsaw University}}

{\centerline {\it Ho\.{z}a 69,~~PL--00--681 Warszawa, ~Poland}}

\vspace{1.0cm}

{\centerline {\bf Abstract}}

\vspace{0.3cm}

 The existence of two sterile neutrinos $\nu_s $ and $\nu'_s $ (blind to all 
Standard--Model interactions) is shown to be implied by a model of fermion 
"texture" that we develop since some time. They may mix nearly maximally with 
two of three conventional neutrinos, say $\nu_e$ and $\nu_\mu$, thus leading to
neutrino oscillations, say $\nu_e \rightarrow \nu_s $ and $\nu_\mu \rightarrow 
\nu'_s $, with nearly maximal amplitudes. Then, they can be responsible for the
observed deficits of solar $\nu_e$'s and atmospheric $\nu_\mu $'s, respectiv%
ely, but by themselves do not help to explain the LSND results for $\nu_\mu 
\rightarrow \nu_e $ oscillations. On the other hand, they are consistent with 
the CHOOZ negative result. At the moment, the experiment cannot decide, whether
the deficit of atmospheric $\nu_\mu $'s, confirmed by the recent \SK findings,
has to be related to the oscillations $\nu_\mu \rightarrow \nu_\tau $ or $
\nu_\mu \rightarrow \nu'_s $. In the last Section of the paper, a new notion of
"non--Abelian spin--1/2 fermions" is presented in the context of a composite 
option for fermion families.

\vspace{0.3cm}

\ni PACS numbers: 12.15.Ff , 12.90.+b , 14.60.Gh

\vspace{2.0cm}

\ni August 1998

\vfill\eject

\pagestyle {plain}

\setcounter{page}{1}

\vspace{0.3cm}

\ni {\bf 1.Introduction}

\vspace{0.3cm}

 The hypothetic sterile neutrinos, by definition interacting only gravitat%
ionally, are blind to all Standard--Model interactions, in contrast to the 
conventional neutrinos (or, rather, their lefthanded parts) which participate 
first of all in the weak sector of Standard--Model interactions. Such Standard%
--Model--inactive fermions are invoked from time to time by theorists, who want
to explain [1] through neutrino oscillations not only the observed deficits of 
solar and atmospheric neutrinos, but also the results of LSND experiment. The 
sterile neutrinos may also form a Standard--Model--inactive fraction of the 
dark matter.

 In the present paper, we demonstrate how two different sterile neutrinos are 
implied by a model of fermion "texture" [2,3] that we develop since some time. 
As shown previously, this model justifies [2] the existence of three and only 
three families of conventional leptons and quarks $(\nu_e\,,\,e^-\,,\,u\,,\,
d)$, $(\nu_\mu\,,\,\mu^-\,,\,c\,,\,s)$, $(\nu_\tau\,,\,\tau^-\,,\,t\,,\,b)$ 
and, moreover, describes reasonably [3] the masses and mixing parameters 
of quarks and charged leptons, making also some useful suggestions as to 
neutrinos. Note that in this model all neutrinos are Dirac particles having 
both lefthanded and righthanded parts.

 In order to make our presentation fairly comprehensible, we will first 
recapitulate briefly the basic features of the model in its part concerning the
existence of fundamental--particle families [2]. Then, we shall discuss 
the existence of two sterile neutrinos and the related neutrino oscillations.

\vspace{0.3cm}

\ni {\bf 2. Dirac's generalized square root}

\vspace{0.3cm}

 The starting point of our model is the conjecture that {\it all} kinds of 
matter's fundamental particles existing in Nature can be deduced from Dirac's
square--root procedure $\sqrt{p^2} \rightarrow \Gamma \cdot p $.

 As is easy to observe, this procedure leads in general to the sequence $ N = 
1,2,3,\ldots $ of different (generally reducible) representations 

\begin{equation}
\Gamma^\mu \equiv {\frac{1}{\sqrt{N}}}\sum_{i=1}^{N} \gamma^\mu_i 
\end{equation}

\ni of the Dirac algebra

\begin{equation}
\left\{ \Gamma^{\mu},\,\Gamma^{\nu} \right\} = 2g^{\mu\nu}\;,
\end{equation}

\ni constructed with the use of the sequence $ N = 1,2,3,\ldots $ of Clifford 
algebras

\begin{equation}
\left\{\gamma^\mu_i,\gamma^\nu_j \right\} = 2\delta_{ij} g^{\mu\nu}\;\;(i,j = 
1,2,\ldots,N)\;.
\end{equation}

\ni Then, the sequence  $ N = 1,2,3,\ldots $ of Dirac--type equations follows,

\begin{equation}
\left\{ \Gamma \cdot \left[ p - gA(x) \right] - M \right\} \psi(x) = 0\;,
\end{equation}

\ni where $ g\Gamma\cdot A(x)$ may symbolize the minimal coupling of $\psi(x)$ 
to the Standard--Model gauge fields $A_\mu (x) $ including all $ SU(3) \times
SU_L(2) \times U(1)$ coupling matrices: $\lambda $'s, $\tau $'s, $ Y $ and 
$\Gamma^5 \equiv i\Gamma^0 \Gamma^1 \Gamma^2 \Gamma^3 $.

 In Eqs. (4) the matrices (1) can be presented in the reduced forms

\begin{equation}
\Gamma^\mu \, = \, \gamma^\mu \otimes 
\underbrace{{\bf 1}\otimes \cdots \otimes {\bf 1}}_{{\ (N-1){\rm times}}}      
\end{equation}

\ni with $\gamma^\mu$ and {\bf 1} denoting the usual $4\times 4$ Dirac 
matrices. Then, the Dirac--type equations (4) can be rewritten as 

\begin{equation}
\left\{\gamma \cdot \left[ p - g A(x) \right] - M \right\}_{\alpha_1\beta_1}
\psi_{\beta_1\alpha_2 \cdots \alpha_N}(x) = 0 
\end{equation}

\ni with $\psi(x) = \left( \psi_{\alpha_1\alpha_2 \cdots \alpha_N}(x)\right)
$, where $ \alpha_1,\,\alpha_2, \cdots,\,\alpha_N $ stand for $ N $ Dirac bi%
spinor indices: $\alpha_i = 1,2,3,4$ for $ i = 1,2,\ldots,N $. Here, the chiral
representations are used to define all $\alpha_i \;\;(i = 1,2,\ldots,N)$. This 
means that $\alpha_i = 1,2,3,4 $ correspond to four different pairs (1,1), 
(1,-1), (-1,1), (-1,-1) of eigenvalues of the matrices

\begin{equation}
\Gamma^5_i \equiv i\Gamma^0_i \Gamma^1_i \Gamma^2_i \Gamma^3_i\;,\;\Sigma_i^3
\equiv i \Gamma^5_i \Gamma^0_i \Gamma^3_i\;,
\end{equation}

\ni simultaneously diagonal for all $ i $, which choice is allowed because all 
$\Gamma^5_i $ and $\Sigma_i^3 $ commute both for equal and different $ i $. The
$\Gamma^\mu_i $ matrices $(i = 1,2,\ldots,N)$ appearing in Eqs. (7) are defined
as $ N $ (properly normalized) Jacobi combinations of $\gamma^\mu_i $ matrices 
$(i = 1,2,\ldots,N)$, where in particular $\Gamma^\mu_1 \equiv \Gamma^\mu $ 
is given as in Eq. (1). Then, $\{\Gamma^\mu_i\,,\,\Gamma^\nu_j\} = 2\delta_{ij}
g^{\mu\nu}$ $(i,j = 1,2,\ldots,N)$ due to Eqs. (3), and also $\{\Gamma^\mu_i\,,
\,\Gamma^5_j\} = 0 $, but $\left[ \Gamma^5_i\,,\,\Gamma^5_j \right] = 0$, where
particularly $\Gamma^5_1 \equiv \Gamma^5 $. Note that in the one--body Dirac--%
type equations (4) there appear only the "centre--of--mass" $\Gamma^\mu_1 $ 
matrices, while all "relative" matrices $\Gamma^\mu_2,\ldots,\Gamma^\mu_N $ are
absent. In spite of this, all $\alpha_1,\alpha_2,\ldots,\alpha_N $ are present 
in Eqs. (6): both the "centre--of--mass" Dirac bispinor index $\alpha_1 $ as 
well as the "relative" Dirac bispinor indices $\alpha_2,\ldots,\alpha_N $, the 
latter are decoupled, however, even in the presence of Standard--Model coupling
$ g\Gamma_i \cdot A(x)$.

 For $ N = 1 $ Eq. (6) is obviously the usual Dirac equation, for $ N = 2 $ it 
is known as the Dirac form [4] of the K\"{a}hler equation [5], whilst for $ N =
3,4,5,\ldots $ we obtain {\it new} Dirac--type equations [2].

 If the Standard--Model coupling $ g\Gamma \cdot A(x)$ is really present in 
Eqs. (6), then the Dirac bispinor index $\alpha_1$, which is the only $\alpha_i
$ affected by the gauge fields $ A_\mu(x) $, is {\it distinguished} by its cor%
relation with the set of all diagonal Standard--Model charges ascribed to any 
particle of the fields $\psi_{\alpha_1\,\alpha_2\ldots\alpha_N}(x) $ (a label 
$ f $ of this set is here suppressed). The remaining Dirac bispinor indices $
\alpha_2,\ldots,\alpha_N $ are all decoupled and so, physically unobservable 
in the gauge fields $ A_\mu(x) $. It is natural to conjecture that they are 
physically {\it undistinguishable} and, therefore, are formal objects obeying 
Fermi statistics along with Pauli principle. This implies that $\psi_{\alpha_1
\,\alpha_2\ldots\alpha_N}(x) $ is fully {\it antisymmetric} with respect to $
\alpha_2,\ldots,\alpha_N $.

 The above conjecture, together with the probabilistic interpretation of wave 
functions $\psi_{\alpha_1\,\alpha_2\ldots\alpha_N}(x) $ and the requirement of 
their relativistic covariance applied to {\it all} bispinor indices $\alpha_1,
\alpha_2,\ldots,\alpha_N $, leads to the conclusion that there are {\it three} 
(and only three) families $ N = 1,3,5 $ of leptons and quarks [2], and {\it two
} (and only two) families $ N = 2,4 $ of some, not yet observed, fundamental
scalars [6]. They correspond to the wave functions

\vspace{-0.2cm}

\begin{eqnarray}
\psi^{(1)}_{\alpha_1} & \equiv & \psi_{\alpha_1}\;, \nonumber \\
\psi^{(3)}_{\alpha_1} & \equiv & \frac{1}{4}\left(C^{-1}\gamma^5 \right)_{
\alpha_2\alpha_3} \psi_{\alpha_1\alpha_2\alpha_3}  = \psi_{\alpha_1
\,12} =  \psi_{\alpha_1\,34}\;, \nonumber \\
\psi^{(5)}_{\alpha_1} & \equiv & \frac{1}{24}\varepsilon_{\alpha_2\alpha_3
\alpha_4\alpha_5}\psi_{\alpha_1\alpha_2\alpha_3\alpha_4\alpha_5} =
\psi_{\alpha_1\,1234} 
\end{eqnarray} 

\vspace{-0.2cm}

\ni and

\vspace{-0.2cm}

\begin{eqnarray}
\phi^{(2)} & \equiv & \frac{1}{2\sqrt{2}}\left(C^{-1}\gamma^5 \right)_{
\alpha_1\alpha_2} \psi_{\alpha_1\alpha_2}  = \frac{1}{\sqrt{2}}\left( 
\psi_{12} - \psi_{21} \right) = \frac{1}{\sqrt{2}}\left( \psi_{34} - 
\psi_{43}\right)\;, \nonumber \\ 
\phi^{(4)} & \equiv & \frac{1}{6\sqrt{4}}\varepsilon_{\alpha_1\alpha_2
\alpha_3\alpha_4}\psi_{\alpha_1\alpha_2\alpha_3\alpha_4} = \frac{1}{
\sqrt{4}}\left( \psi_{1234} - \psi_{2134} + \psi_{3412} - 
\psi_{4312} \right)\;,
\end{eqnarray} 

\ni respectively. Each of these wave functions carries the (here suppressed) 
Standard--Model label $ f = \nu\,,\,e\,,\,u\,,\,d $ denoting four sorts of 
fundamental particles corresponding to the signature of conventional neutrinos 
$\nu $ and charged leptons $ e $ as well as up quarks $ u $ and down quarks $ 
d $, all four following from the Standard Model (though the existence of three 
and two fundamental--particle families does not follow from it). In the case of
fundamental fermions, the three families are, of course, $(\nu_e\,,\,e^-\,,\,u
\,,\,d)$, $(\nu_\mu\,,\,\mu^-\,,\,c\,,\,s)$, $(\nu_\tau\,,\,\tau^-\,,\,t\,,\,b)
$, while in the case of fundamental scalars one of ({\it a priori}) possible 
options may be that the two families correspond to the first and second fermion
family [6].

 Now, in contrast, if the Standard--Model coupling $ g \Gamma \cdot A(x) $ is 
absent from Eqs. (6), then {\it only} physically undistinguishable {\it i.e.}, 
antisymmetric bispinor indices $\alpha_i\;\;(i = 1,2,\ldots,N)$ can appear at 
the wave functions $\psi_{\alpha_1\alpha_2\ldots\alpha_N}(x)$. In this case, 
the argument similar to the used before shows that on the fundamental level 
there are {\it two} (and only two) Standard--Model--inactive spin--1/2 fermions
$ N = 1,3 $ corresponding to the wave functions

\vspace{-0.3cm}

\begin{eqnarray}
\psi^{(1)}_{\alpha_1} & \equiv & \psi_{\alpha_1}\;, 
\nonumber \\ \psi^{(3)}_{\alpha_1} & \equiv & \frac{1}{6}
\left(C^{-1}\gamma^5 \right)_{\alpha_1\,\alpha_2}\varepsilon_{\alpha_2\alpha_3
\alpha_4\alpha_5}\psi_{\alpha_3\alpha_4\alpha_5} 
\end{eqnarray} 

\ni (with no suppressed $ f $ label). They can be identified with two {\it 
sterile neutrinos} denoted in this paper by $\nu_s $ and $\nu'_s $, respect%
ively. Analogically, on the fundamental level there should exist also {\it two}
(and only two) Standard--Model--inactive spin--0 bosons $ N = 2,4 $ that may be
called {\it sterile scalars}, $\phi^{(2)}$ and $\phi^{(4)}$ (with no suppressed
$ f $ label).

\vspace{0.3cm}

\ni {\bf 3. Neutrino oscillations involving $\nu_s $ and $\nu'_s $}

\vspace{0.3cm}

 Let us conjecture tentatively that the sterile neutrinos $\nu_s $ and $\nu'_s
$ are compelled to mix nearly maximally with the conventional neutrinos $\nu_e$
and $\nu_\mu $, respectively, in order to form four related neutrino mass 
states $\nu_1 $ or $\nu_4 $ and $\nu_2$ or $\nu_5 $. Other neutrino mixings are
assumed not to appear at all or to be negligible. In particular, the third con%
ventional neutrino $\nu_\tau $ is left not mixed and so, $\nu_3 = \nu_\tau$ is 
a neutrino mass state. Evidently, the mixings of $\nu_e $ with $\nu_s $ and $
\nu_\mu $ with $\nu'_s $ would be forbidden, if the electroweak $ SU_L(2)\times
U(1) $ symmetry were not spontaneously broken. Thus, we can say that neutrino 
oscillations, being a consequence of these mixings, are caused in fact by the 
spontaneous breaking of electroweak symmetry (if, of course, sterile neutrinos 
exist).

 Under the above conjecture, the unitary transformation $\nu_I  = \sum_\alpha
V_{I\,\alpha} \nu_\alpha $ between neutrino mass states $\nu_I = \nu_1,\,\nu_2,
\,\nu_3,\,\nu_4,\,\nu_5 $ and neutrino flavor states $\nu_\alpha = \nu_e,\,
\nu_\mu,\,\nu_\tau,\,\nu_s,\,\nu'_s $ is given as

\vspace{-0.2cm}

\begin{eqnarray}
\nu_1 = V_{11}\nu_e + V_{14}\nu_s & , & \nu_4 = V_{41}\nu_e + V_{44}\nu_s \;, 
\nonumber \\
\nu_2 = V_{22}\nu_\mu + V_{25}\nu'_s  & , & \nu_5 = V_{52}\nu_\mu + V_{55}
\nu'_s \;, \nonumber \\
\nu_3 & = & \nu_\tau \;,
\end{eqnarray}

\ni where the nonzero coefficients are

\vspace{-0.2cm}

\begin{eqnarray}
V_{11} = V_{44} = \frac{1}{\sqrt{1+Y^2}}\; &,& V_{14} = - V_{41}^* = -\frac{Y
}{\sqrt{1+Y^2}} e^{i\varphi}\;, \nonumber \\ V_{22} = V_{55} = \frac{1}{\sqrt{1
+ X^2}} \; &,& V_{25} = -V^*_{52} = -\frac{X}{\sqrt{1+X^2}} e^{i\varphi'} \;,
\nonumber \\ V_{33} & = & 1 
\end{eqnarray}

\ni (in Eqs. (11) and (12), for notation convenience, we write $ V_{IJ}$ in 
place of $ V_{I\,\alpha}$, where $I,J = 1,2,3,4,5$). The magnitudes of these 
coefficients are determined by the parameters

\vspace{-0.2cm}

\begin{eqnarray}
Y & = & \frac{M_{11} - m_{\nu_1}}{|M_{14}|} = - \frac{M_{44} - m_{\nu_4}}{|M_{
14}|}\;, \nonumber \\ & & \nonumber \\
X & = & \frac{M_{22} - m_{\nu_2}}{|M_{25}|} = - \frac{M_{55} - m_{\nu_5}}{|M_{
25}|}
\end{eqnarray}

\ni involving neutrino masses

\vspace{-0.2cm}

\begin{eqnarray}
m_{\nu_1,\,\nu_4} & = & \frac{M_{11}+M_{44}}{2} \mp \sqrt{\left(\frac{M_{11} - 
M_{44}}{2}\right)^2 + |M_{14}|^2} \;, \nonumber \\ 
m_{\nu_2,\,\nu_5} & = & \frac{M_{22}+M_{55}}{2} \mp \sqrt{\left(\frac{M_{22} - 
M_{55}}{2}\right)^2 + |M_{25}|^2}\;.
\end{eqnarray}

\ni On the other hand $m_{\nu_3} = M_{33} $. Here, $ \left( M_{IJ}\right)\;\;
(I,J = 1,2,3,4,5) $ is a $ 5\times 5 $ neutrino mass matrix with $ M_{14} = 
M^*_{41} = |M_{14}| \exp i\varphi $ and $ M_{25} = M^*_{52} = |M_{25}| \exp 
i\varphi' $ as the only off-diagonal elements. Then, $ \left( V_{IJ}\right)\;\;
(I,J = 1,2,3,4,5) $ is a $ 5\times 5$ lepton counterpart of the familiar \CKM 
matrix for quarks, where now $ V_{14} = -V^*_{41}$ and $ V_{25} = -V^*_{52}$ 
are the only nonzero off-diagonal elements.

 Some (here neglected) small corrections to the neutrino mixings (11) may be 
caused by possible small deviations of the charged--lepton mass matrix from a 
diagonal form [3]. In fact, these deviations produce small deviations of 
the related diagonalizing unitary matrix from the unit matrix. In turn, such 
a charged--lepton diagonalizing matrix contributes multiplicatively to the 
lepton \CKM matrix [3], changing a little its leading form (12) (in particular,
nearly all zero elements of its leading form become nonzero but small).

 Now, making use of~Eqs. (12), we can calculate the probabilities of neutrino 
oscil\-lations $\nu_e \rightarrow \nu_s $ and $\nu_\mu \rightarrow \nu'_s $ 
(in 
the vacuum) from the general formula

\vspace{-0.1cm}

\begin{equation}
P(\nu_\alpha \rightarrow \nu_\beta) = |\langle\nu_\beta |\nu_\alpha(t)\rangle
|^2 = \sum_{K\,L}V_{L\,\beta}V^*_{L\,\alpha}V^*_{K\,\beta}V_{K\,\alpha} 
\exp\left(i\frac{\Delta m^2_{LK}}{2|\vec{p}|}\,t\right)\;,
\end{equation}

\ni where $\Delta m^2_{LK} = m^2_{\nu_L} - m^2_{\nu_K} $ (on the rhs of Eq.
(15), for notation convenience, we will replace $\alpha\,,\,\beta $ by $I\,,\,J
= 1,2,3,4,5 $). Here, $\nu_\alpha(0) = \nu_\alpha$, $\langle\nu_\beta| = 
\langle 0| \nu_\beta $, $\langle\nu_\beta |\nu_\alpha \rangle = \delta_{\beta
\,\alpha} $ and, as usual, $t/|\vec{p}| = L/E\,\;( c = 1 = \hbar)$, what is 
equal to $ 4\times 1.2663 L/E $ if $\Delta m^2_{LK} \;,\; L $ and $ E $ are
measured in eV$^2$, m and MeV, respectively ($ L $ is, of course, the source%
--detector distance). Further on, we will denote

\vspace{-0.1cm}

\begin{equation}
x_{LK} = 1.2663 \frac{\Delta m^2_{LK} L}{E}  
\end{equation} 

\ni and use the identity $ \cos 2x_{LK} = 1 - 2\sin^2 x_{LK}$.

 In such a way, we derive the following formulae for probabilities of neutrino 
oscillations $\nu_e \rightarrow \nu_s $ and $\nu_\mu \rightarrow \nu'_s $ (in 
the vacuum):

\vspace{-0.2cm}

\begin{eqnarray}
P\left(\nu_e \rightarrow \nu_s\right) & = & 4 \frac{Y^2}{(1+Y^2)^2} 
\sin^2 x_{41} \;, \nonumber \\
P\left(\nu_\mu \rightarrow \nu'_s\right) & = & 4\frac{X^2}{(1+X^2)^2}
\sin^2 x_{52} \;,
\end{eqnarray}

\ni while all other $P\left(\nu_\alpha \rightarrow \nu_\beta\right)$ with $
\alpha \neq \beta $ vanish [except, of course, for $P\left(\nu_s \rightarrow 
\nu_e \right) = P\left(\nu_e \rightarrow \nu_s\right)$ and $P\left(\nu'_s 
\rightarrow \nu_\mu\right) = P\left(\nu_\mu \rightarrow \nu'_s \right)$]. Thus,
the neutrino--oscillation formulae (in the vacuum) for survival 
probabilities of
$\nu_e $ and $\nu_\mu $ are

\vspace{-0.2cm}

\begin{eqnarray}
P\left(\nu_e \rightarrow \nu_e\right) & = & 1 - 4 \frac{Y^2}{(1+Y^2)^2} 
\sin^2 x_{41}\;, \nonumber \\ 
P\left(\nu_\mu \rightarrow \nu_\mu\right) & = & 1 - 4 \frac{X^2}{(1+X^2)^2} 
\sin^2 x_{52}\;.
\end{eqnarray}

 In the case of solar neutrinos, the observed deficit of $\nu_e $'s can be 
explained through the neutrino oscillations (in the vacuum), when using the 
two--flavor formula for survival probability of $\nu_e $,

\begin{equation}
P\left(\nu_e \rightarrow \nu_e\right) = 1 - \sin^2 2\theta_{\rm solar}
\sin^2 \left(1.27 \frac{\Delta m^2_{\rm solar}\, L}{E} \right)\;.
\end{equation}

\ni with the parameters [7]

\begin{equation}
\sin^2 2\theta_{\rm solar} \sim 0.65\;\;{\rm to}\;\;1\;\;,\;\;\Delta m^2_{\rm 
solar} \sim (5\;\;{\rm to}\;\;8) \times 10^{-11}\;{\rm eV}^2 \;.
\end{equation}

\ni These give the so called vacuum fit, in contrast to two other known fits 
based on the resonant MSW mechanism [8] in the Sun matter. In our model of 
neutrino oscillations (where $\nu_e \rightarrow \nu_s$ oscillations are 
responsible for the deficit of solar $\nu_e $'s), this fit leads to

\begin{equation}
\frac{4 Y^2}{(1+Y^2)^2} \sim 0.65\;\;{\rm to}\;\; 1\;\;,\;\;\Delta m^2_{41} 
\sim (5\;\;{\rm to}\;\;8) \times 10^{-11}\;{\rm eV}^2 
\end{equation}

\ni (as $m^2_{\nu_4} > m^2_{\nu_1}$). Hence, $ Y \sim 0.507 $ to 1 and so, we 
get a large mixing of $\nu_e $ with $\nu_s $: $ V_{11} = V_{44} \sim 0.892 $ 
to $1/\sqrt{2}$ and $ V_{14} = - V_{41}^* \sim -(0.452\;\;{\rm to}\;\;1/\sqrt{
2}) \exp i\varphi $ (the phase $\varphi $ remains not determined).

In the case of atmospheric neutrinos, the recent findings of the \SK experiment
[9] show that the observed deficit of $\nu_\mu $'s can be explained also 
through the neutrino oscillations (in the vacuum), when making use of the two%
--flavor formula for survival probability of $\nu_\mu $,

\begin{equation}
P\left(\nu_\mu \rightarrow \nu_\mu \right) = 1 - \sin^2 2\theta_{\rm atm}
\sin^2 \left(1.27 \frac{\Delta m^2_{\rm atm} L}{E} \right)\;,
\end{equation}

\ni with the parameters 

\begin{equation}
\sin^2 2\theta_{\rm atm} \sim 0.82\;\;{\rm to}\;\;1\;\;,\;\;\Delta m^2_{\rm 
atm} \sim (0.5\;\;{\rm to}\;\;6) \times 10^{-3}\;{\rm eV}^2 \;.
\end{equation}

\ni In our model of neutrino oscillations (where $\nu_\mu \rightarrow \nu'_s$ 
oscillations are responsible for the deficit of atmospheric $\nu_\mu $'s), this
implies 

\begin{equation} 
\frac{4 X^2}{(1+X^2)^2} \sim 0.82\;\;{\rm to}\;\;1\;\;,\;\;\Delta m^2_{52}
\sim (0.5\;\;{\rm to}\;\;6) \times 10^{-3}\;{\rm eV}^2 
\end{equation}

\ni (as $m^2_{\nu_5} > m^2_{\nu_2}$). Hence, $ X \sim 0.636 $ to 1 and thus, 
we obtain a 
large mixing of $\nu_\mu $ with $\nu'_s $: $ V_{22} = V_{55} \sim 0.844 $ to $
1/\sqrt{2}$ and $ V_{25} = - V_{52}^* \sim -(0.537\;\;{\rm to}\;\;1/\sqrt{2})
\exp i\varphi' $ (the phase $\varphi' $ remains not determined).

 On the other hand, the CHOOZ experiment [10] found no evidence for neutrino--%
oscillation modes of $\bar{\nu}_e $ in a parameter region overlapping the range
(23) of $\sin^2 2\theta_{\rm atm}$ and $\Delta m^2_{\rm atm} $, what shows that
within this parameter range there are no mixings of $\nu_e $ with $\nu_\mu $, 
$\nu_\tau $, $\nu_s $, $\nu'_s$. In particular for $\nu_\mu$, this is 
consistent with the assumed dominance of mixing between $\nu_\mu$ and $\nu'_s
$ over mixing between $\nu_\mu$ and $\nu_e $ within the range (23) of $\sin^2 
2\theta_{\rm atm}$ and $\Delta m^2_{\rm atm} $ (at the moment, however, it can%
not be decided experimentally [9], whether the mixing of $\nu_\mu$ with $\nu'_s
$ or the here neglected mixing of $\nu_\mu$ with $\nu_\tau$ is responsible for
the deficit of atmospheric $\nu_\mu$'s). For $\nu_s $, this requires that the 
assumed strong mixing of $\nu_e $ with $\nu_s $ must correspond to parameters 
$\sin^2 2\theta_{\rm solar}$ and $\Delta m^2_{\rm solar}$ belonging to a range 
very different from (23) [in fact, they can lie in the range (20)]. Finally for
$\nu_\tau $, the lack of mixing between $\nu_\tau $ and $\nu_e $ is one of 
necessary and sufficient conditions for the assumed identity $\nu_3 = \nu_\tau 
$ (another is the lack of mixing between $\nu_\tau$ and $\nu_\mu $, if this 
really is true).

 Of course, the sterile neutrinos $\nu_s $ and $\nu'_s $ by themselves cannot 
help to explain the results of LSND experiment [11] which gave evidence for 
$\bar{\nu}_\mu \rightarrow \bar{\nu}_e $ and  $\nu_\mu \rightarrow \nu_e $ 
oscillations corresponding to a much larger $\Delta m^2_{\rm LSND}$ than both 
$\Delta m^2_{\rm solar}$ and $\Delta m^2_{\rm atm}$. These oscillations, if
eventually confirmed, would require a considerable mixing of $\nu_\mu $ with 
$\nu_e $, corresponding to parameters $\sin^2 2\theta_{\rm LSND}$ and $\Delta 
m^2_{\rm LSND} $ lying in a range very different from (23). This mixing should 
be stronger than that induced by the (mentioned before) nondiagonal charged--%
lepton corrections appearing in our model of fermion "texture" [3].

 The comparison of mass squared differences $\Delta m^2_{41}$ and $\Delta 
m^2_{52}$ as estimated in Eqs. (21) and (24) suggests that $ m^2_{\nu_1}$
and $ m^2_{\nu_4}$ are possibly much smaller than $ m^2_{\nu_2}$ and $ m^2_{
\nu_5}$ (alternatively, $ m^2_{\nu_1}$ and $ m^2_{\nu_4}$ may be much more 
degenerate than $ m^2_{\nu_2}$ and $ m^2_{\nu_5}$).

\vspace{0.3cm}

\ni {\bf 4. Outlook: Non--Abelian spin--1/2 fermions}

\vspace{0.3cm}

 When the Dirac--type equations (4) are considered, one may ask a (perhaps) 
profound question, as to whether these one--body equations could be understood 
physically as point--like limits of some $ N $--body equations for tight bound 
states of $ N $ spin--1/2 preons with equal masses. If it was so, the four--%
positions of such subelementary constituents (called here preons, as usual)
should tend practically (within the bound states) to their centre--of--mass 
four--position,

\begin{equation}
x_i = X + \delta x_i \rightarrow X\;,\; X \equiv \frac{1}{N}\sum_{i=1}^{N} 
x_i \;,\;\sum_{i=1}^{N} \delta x_i \equiv 0 \;,
\end{equation}

\ni while then $ \delta p_i $, defined by their four--momenta  

\begin{equation}
p_i = P + \delta p_i \;,\; P \equiv \sum_{i=1}^{N} 
p_i \;,\;\sum_{i=1}^{N} \delta p_i \equiv 0 \;,
\end{equation}

\ni should vanish in action on the wave functions [here, $ x_i = (t_i
\,,\,\vec{x}_i) $, $ \delta x_i = (\delta t_i\,,\,\delta \vec{x}_i)$ and $ X = 
(t, \vec{X})$].

 Of course, the physical mechanism for realization of such practically point--%
like limits in $ N $--body systems would be provided by an unknown, very strong
and shortrange attraction between their $ N $ constituents (described, for 
convenience, in the equal--time formalism, where $\delta t_i \equiv 0 $ and 
$\delta p_i^0 $ vanish in action on the wave functions). The (necessarily) 
non--Standard--Model nature of such an attraction would be certainly the most 
obscure aspect of the compound option for the Dirac--type equations (4).

 Let us denote by $ P_i $ and $ X_i $ $(i = 1,2,\ldots,N)$, with $ P_1 \equiv P
$ and $ X_1 \equiv X $, the (properly normalized) Jacobi combinations of 
four--momenta $ p_i $ and four--positions $ x_i $ $ (i = 1,2,\ldots,N) $, 
respectively, for $ N $ particles. Then,
   
\begin{equation}
\left[ P_i^\mu\,,\,X_j^\nu \right] = i \delta_{ij} g^{\mu\nu}\;.
\end{equation}

\ni Making use of this notation, we can write the identities

\begin{equation}
\sum_{i=1}^{N} \left(\gamma_i \cdot p_i - m_i \right) = \frac{1}{\sqrt{N}}
\sum_{i=1}^{N} \left(\Gamma_i \cdot P_i - \sqrt{N} m_i \right)\;,
\end{equation}

\ni where $\Gamma^\mu_i $ $(i = 1,2,\ldots,N)$, with $\Gamma^\mu_1 \equiv 
\Gamma^\mu $, stand for the (properly normalized) Jacobi combinations of 
$\gamma^\mu_i $ matrices $(i = 1,2,\ldots,N)$ for $ N $ particles [the $
\Gamma^\mu_i $ matrices were already introduced in Eqs. (7), though only in 
reference to the one--body Dirac--type equations(4)]. Then,

\begin{equation}
\left\{\Gamma^\mu_i,\Gamma^\nu_j \right\} = 2\delta_{ij} g^{\mu\nu}\;\;(i,j = 
1,2,\ldots,N)\;,
\end{equation}

\ni as follows from Eqs. (3). Here, in particular, $\Gamma^\mu_1 \equiv 
\Gamma^\mu $ is given as in Eq. (1).

 In this notation, the natural candidates for the hypothetic $ N $--body 
equations would be

\begin{equation}
\left\{\Gamma_1 \cdot \left[P_1 - g A(X_1) \right] + \sum_{i=2}^N \Gamma_i
\cdot P_i - \sqrt{N}\left(\sum_{i=1}^N m_i + I \right) \right\}\psi(X_1,X_2,
\ldots,X_N) = 0\;,
\end{equation}

\ni where  $ I(X_2,\ldots,X_N) $ would symbolize the unknown non--Standard--%
Model attraction between $ N $ constituents. In Eqs. (30), the Standard--Model 
gauge fields $ A_\mu(x) $ are coupled to the hypothetic $ N $--body systems 
at four--points describing their centre--of--mass four--positions $ X_1 \equiv 
X $. This is approximately true, when $ A_\mu(X + \delta x_i) $ are only weakly
dependent on $\delta x_i $.

 In the point--like limits, where the relative four--positions $ X_2,\ldots,X_N
$ ({\it i.e.,} also all $\delta x_i$) tend to zero and then the relative 
four--%
momenta $P_2,\ldots,P_N $ ({\it i.e.,} also all $\delta p_i$) vanish in action 
on the wave functions, Eqs. (30) are really reduced to the Dirac--type 
equations (4) with $ p \equiv P \equiv P_1 $, $ X \equiv X \equiv X_1 $, 
$\Gamma \equiv \Gamma_1 $ and $ M \equiv \sqrt{N}\,(N m + I_{X_i\rightarrow 0})
$  ($ m_i \equiv m $). Note that $ M $ grows with $ N $ faster than linearly.
   
 In the equal--time formalism, where the relative times $t_2,\ldots,t_N $ ({\it
i.e.,} also all $\delta t_i$) are zero and the relative energies $P^0_2,\ldots
,P^0_N $ ({\it i.e.,} also all $\delta p^0_i$) vanish in action on the wave 
functions, Eqs. (30) assume the forms

\begin{eqnarray}
P^0_1\psi(\vec{X}_1,\vec{X}_2,\ldots,\vec{X}_N,t) & = & \left\{\Gamma^0_1
\vec{\Gamma_1} \cdot\left[\vec{P}_1 - g \vec{A}(\vec{X_1},t) \right] +  g 
A^0(\vec{X}_1,t) + \sum_{i=2}^N \Gamma^0_1\vec{\Gamma_i}\cdot \vec{P}_i\right. 
\nonumber \\
& & \left.\, +\Gamma^0_1\left(\sqrt{N} \sum_{i=1}^N m_i + I_{X^0_i = 0} 
\right)\right\}\psi(\vec{X}_1,\vec{X}_2,\ldots,\vec{X}_N,t) \;,\nonumber \\
\end{eqnarray}

\ni where $ P_1^0 \equiv P^0 = i \partial/\partial t $ and $I_{X^0_i = 0
} = I(\vec{X}_2,\ldots,\vec{X}_N)$. 

 In the point--like limits, Eqs. (31) are reduced to the equations

\begin{equation}
p^0 \psi(\vec{x},t) = \left\{ \Gamma^0 \vec{\Gamma}\cdot \left[\vec{p} - 
g \vec{A}(\vec{x},t)\right] + g A^0(\vec{x},t) + \Gamma^0 M\right\}
\psi(\vec{x},t)
\end{equation}

\ni with $p \equiv P \equiv P_1\,,\; x \equiv X \equiv X_1\,,\; \Gamma \equiv 
\Gamma_1 $ and $ M \equiv \sqrt{N}(N m + I_{X_i\rightarrow 0}) $ ($ m_i \equiv 
m $). Of course, $ p^0 \equiv P^0 = i \partial/\partial t $ and $I_{X_i
\rightarrow 0} $ stands for a reasonably defined point--like limit of $ I $.

 Note that the eigenvalues $ (P^0_1)_{\rm kin}$ of the kinetic part of the 
hamiltonian appearing on the rhs of the state equation (31) get for any $ N $
the form $\pm\left[\vec{P}^2_1 + \ldots + \vec{P}^2_N + N(N m)^2\right]^{1/2}$
$ = \pm\sqrt{N}\left[\vec{p}^{\,2}_1 + \ldots + \vec{p}^{\,2}_N + N(N m)^2
\right]^{1/2}$, as if our $ N $--body system were a single Dirac particle with 
the mass $ N m $ in a $(3N+1)$--dimensional spacetime (notice, however, the 
additional factor $\sqrt{N}$).

 A fundamental feature of Eqs. (30) is that, {\it via} $\Gamma^\mu_i$ ($i = 1,
2, \ldots,N$), they contain $ N $ Dirac nonconventional $\gamma^\mu_i$ matrices
($i = 1, 2, \ldots,N$) which do not commute for different particles, in 
contrast to Dirac conventional gammas commuting for different particles [in
fact, the nonconventional $\gamma^\mu_i$ ($i = 1, 2, \ldots,N $) anticommute 
for different $ i $, as is seen from Eqs. (3)]. The spin--1/2 fermions $ i = 
1,2,\ldots,N $ described within an $ N $--body system with the use of such 
nonconventional $\gamma^\mu_i$ matrices ($i = 1, 2, \ldots,N $), anticommuting
for different particles, might be called {\it non--Abelian} spin--1/2 fermions
[12]. In contrast, in the familiar case of Dirac conventional gammas, commuting
for different particles, one could use the term {\it Abelian} spin--1/2 
fermions.
 
 Now, let us observe that the form of spin tensors for spin--1/2 fermions $ i =
1,2,\ldots,N $ is identical in the non--Abelian and Abelian case:  

\begin{equation}
\sigma^{\mu\nu}_i \equiv {\frac{i}{2}}\left[ \gamma^\mu_i\, ,\,\gamma^\nu_i 
\right]
= \left\{ \begin{array}{ll} i\alpha^l_i & {\rm for~}\mu = 0\, ,\, \nu = l\\
\varepsilon^{klm}\sigma^m_i & {\rm for~} \mu = k \, , \, \nu = l
\end{array} \right. \;,
\end{equation}

\ni where $\alpha^l_i \equiv \gamma^0_i\gamma^l_i $ and $\sigma^m_i \equiv 
\gamma^5_i\gamma^0_i\gamma^m_i \equiv \gamma^5_i\alpha^m_i $ with $\gamma^5_i
\equiv i\gamma^0_i\gamma^1_i\gamma^2_i\gamma^3_i $. In fact, for each $ i $ 
the components $\frac{1}{2}\sigma^{\mu\nu}_i $ satisfy in both cases the usual 
Lorentz--group commutation relations, while for different $ i $ they commute 
in both cases as being bilinear in $\gamma^\mu_i $. Also $\gamma^5_i $ commute 
for different $i $ in both cases. So, the total generators of Lorentz group for
a system of $ N $ spin--1/2 fermions are given in both cases by the operators

\begin{equation}
J^{\mu} =  \sum_{i=1}^N \left( x^\mu_i p^\nu_i - x^\nu_i p^\mu_i +\frac{1}{2}
\sigma ^{\mu\nu}_i \right)\;.
\end{equation}

 Let us note, by the way, the following identity valid for the total spin 
tensor in both cases

\begin{equation}
\sum_{i=1}^N \sigma^{\mu\nu}_i = \sum_{i=1}^N \Sigma^{\mu\nu}_i\;,
\end{equation}

\ni where

\begin{equation}
\Sigma^{\mu\nu}_i \equiv \frac{i}{2}\left[\Gamma_i^\mu , \Gamma_i^\nu\right] =
\left\{ \begin{array}{llll} i\,A^l_i & {\rm for} & \mu = 0 , & \nu = l \\
\varepsilon^{klm}\Sigma^m_i & {\rm for} & \mu = k , & \nu = l \end{array} 
\right. 
\end{equation}

\ni with $ A_i^l \equiv \Gamma_i^0\Gamma_i^l$ and $\Sigma^m_i \equiv \Gamma_i^5
\Gamma_i^0\Gamma_i^m \equiv \Gamma_i^5 A^m_i $. Evidently, $\Sigma^{\mu\nu}_1 
\equiv \Sigma^{\mu\nu}$ with $\Sigma^{\mu\nu} \equiv \frac{i}{2}\left[\Gamma^%
\mu,\Gamma^\nu\right] $ is the centre--of--mass spin tensor for the system of 
$ N $ spin--1/2 fermions, while $\Sigma^{\mu\nu}_2,\ldots,\Sigma^{\mu\nu}_N $ 
are its relative spin tensors. All spin tensors $\Sigma^{\mu\nu}_i $, being 
bilinear in $\Gamma^\mu_i $, commute for different $ i $ in both cases.
 
 Thus, in this Section, we can draw the important conclusion that for a system 
of $ N $ spin--1/2 fermions the Lorentz--group commutation relations get {\it 
two} (and only two) realizations: {\it either} with the use of Dirac convent%
ional gammas commuting for different particles, {\it or} with the use of Dirac 
nonconventional gammas anticommuting for different particles. Such an 
intriguing statement seems to support the logical consistency and unexpected 
naturalness of the notion of non--Abelian spin--1/2 fermions. They may provide 
an unconventional alternative for familiar Abelian spin--1/2 fermions in the 
potential structure of particle theory. In this Section, their role as 
hypothetic preons was underlined.

 Finally, we should like to emphasize some unconventional features of the 
quantization procedure which would work in the case of non--Abelian spin--1/2 
fermions. It is not difficult to observe that in the case of spin--1/2 
fermions, {\it only} the Fock--space states related to Dirac conventional 
gammas (commuting for different particles) can be constructed by means of the 
familiar second--quantization procedure based on Fermi creation and annihil%
ation operators for single particles. This is so, because the repetition of 
single--particle creation operators can lead to Fock--space states of 
particles with commuting Dirac gammas {\it only}. In order to construct the 
Fock--space states related to Dirac nonconventional gammas (anticommuting for 
different particles), {\it new} Fermi or Bose operators creating and 
annihilating at once {\it whole} $ N $--particle configurations with odd or 
even $ N = 1,2,3,\ldots $, respectively, must be introduced. Of course, these
$ N $ particles are then non--Abelian spin--1/2 fermions. Such a new procedure 
might be called the {\it third quantization} [12].

\vfill\eject

~~~~
\vspace{0.5cm}

{\bf References}

\vspace{1.0cm}

{\everypar={\hangindent=0.5truecm}
\parindent=0pt\frenchspacing

~1.~For a review {\it cf. e.g.} A.Y. Smirnow in {\it Proc. 28th Inter.
Conf. on High Energy Physics, Warsaw 1996}, eds. Z. Ajduk and A.K. 
Wr\'{o}blewski, World Scientific, 1997.

~2.~W.~Kr\'{o}likowski, {\it Acta Phys. Pol.} {\bf B 21}, 871 (1990); {\it 
Phys. Rev.} {\bf D 45}, 3222 (1992); in {\it Spinors, Twistors, Clifford 
Algebras and Quantum Deformations (Proc. 2nd Max Born Symposium 1992)}, eds. 
Z.~Oziewicz {\it et al.}, Kluwer, 1993. 

~3.~W.~Kr\'{o}likowski, {\it Acta Phys. Pol.} {\bf B 27}, 2121 (1996); {\bf B 
28}, 1643 (1997); {\bf B 29}, 629 (1998); {\bf B 29}, 755 (1998); 
hep--ph/9803323; hep--ph/9808207.

~4.~T.~Banks,~Y.~Dothan and D.~Horn, {\it Phys. Lett.} {\bf B117}, 413 (1982).

~5.~E.~K\"{a}hler, {\it Rendiconti di Matematica} {\bf 21}, 425 (1962); {\it 
cf.}  also D.~Ivanenko and L.~Landau, {\it Z. Phys.} {\bf 48}, 341 (1928).

~6.~W.~Kr\'{o}likowski, {\it Phys. Rev.} {\bf D 46}, 5188 (1992); {\it Acta 
Phys. Pol.} {\bf B 24}, 1149 (1993); {\bf B 26}, 1217 (1995); {\it Nuovo 
Cimento }, {\bf 107 A}, 69 (1994).

~7.~N.~Hata and P.~Langacker, hep--ph/9705339; {\it cf.} also G.L.~Fogli, 
E.~Lisi and D.~Montanino, hep--ph /9709473.

~8.~L.~Wolfenstein, {\it Phys. Rev.} {\bf D 17}, 2369 (1978); S.P. Mikheyev 
and A.~Y.~Smirnow, {\it Sov. J. Nucl. Phys.} {\bf 42}, 913 (1985); {\it 
Nuovo Cimento}, {\bf C 9}, 17 (1986).

~9.~Y. Fukuda {\it et al.} (Super--Kamiokande Collaboration), "Evidence for 
oscillation of atmospheric neutrinos", submitted to {\it Phys. Rev. Lett.};
and references therein.

10.~M. Appolonio {\it et al.} (CHOOZ Collaboration), {\it Phys. Lett.} {\bf B 
420}, 397 (1998).

11.~C.~Athanassopoulos {\it et al.} (LSND Collaboration), {\it Phys. Rev.} 
{\bf C 54}, 2685 (1996); {\it Phys. Rev. Lett.} {\bf 77}, 3082 (1996);
nucl--ex/9709006.

12.~W.~Kr\'{o}likowski, {\it Acta Phys. Pol.} {\bf B 22}, 613 (1991) [E: {\it 
ibid.} {\bf B 23}, 83 (1992)].

\vfill\eject

\end{document}